\documentclass{article}
\usepackage{graphicx}

\begin{document}
\title{X-ray galaxies in nearby filaments}
\author{Anatoliy Tugay\\Dept. of Astronomy \& Space Physics, Faculty of Physics, \\ Taras Shevchenko Kyiv National University, \\ Glushkova Ave. 4, Kyiv, Ukraine \\email: {\tt tugay.anatoliy@gmail.com}}
\date{}
\maketitle
. \\[1mm]
New sample of X-ray galaxies selected from 2XMMi catalogue in SDSS region is analysed in this work. Spatial distribution and X-ray AGN spectral properties are discussed. A new method for extragalactic filament detection and description is proposed. 

{Keywords: X-rays: galaxies; large-scale structure of Universe}

\section{Introduction}

Xgal - a sample of 5021 X-ray emitting galaxies was compiled in \cite{tugay12} based on 2XMM catalog (\cite{watson09}) and HyperLeda database. In \cite{tugay11} it was shown that extragalactic X-ray sources are AGNs and clusters. It was also shown in \cite{tugay13} that the main type of X-ray galaxies is Seyfert 1. X-ray spectra of 42 Seyfert 1 galaxies from SDSS region were analysed in \cite{tugay13a}. These galaxies has radial velocities from 4000 to 39000 km/s (appropriate for filament detection) and does not necessary belongs to known filaments from \cite{smith12}. An example of 100 Mpc-thick slice with Xgal objects and filaments is presented at Fig. 2. For 7 galaxies X-ray spectra were obtained in \cite{tugay13a} at first time. One of them is presented at Fig. 3.

\begin{figure}[b]
\begin{center}
 \includegraphics[width=3.4in]{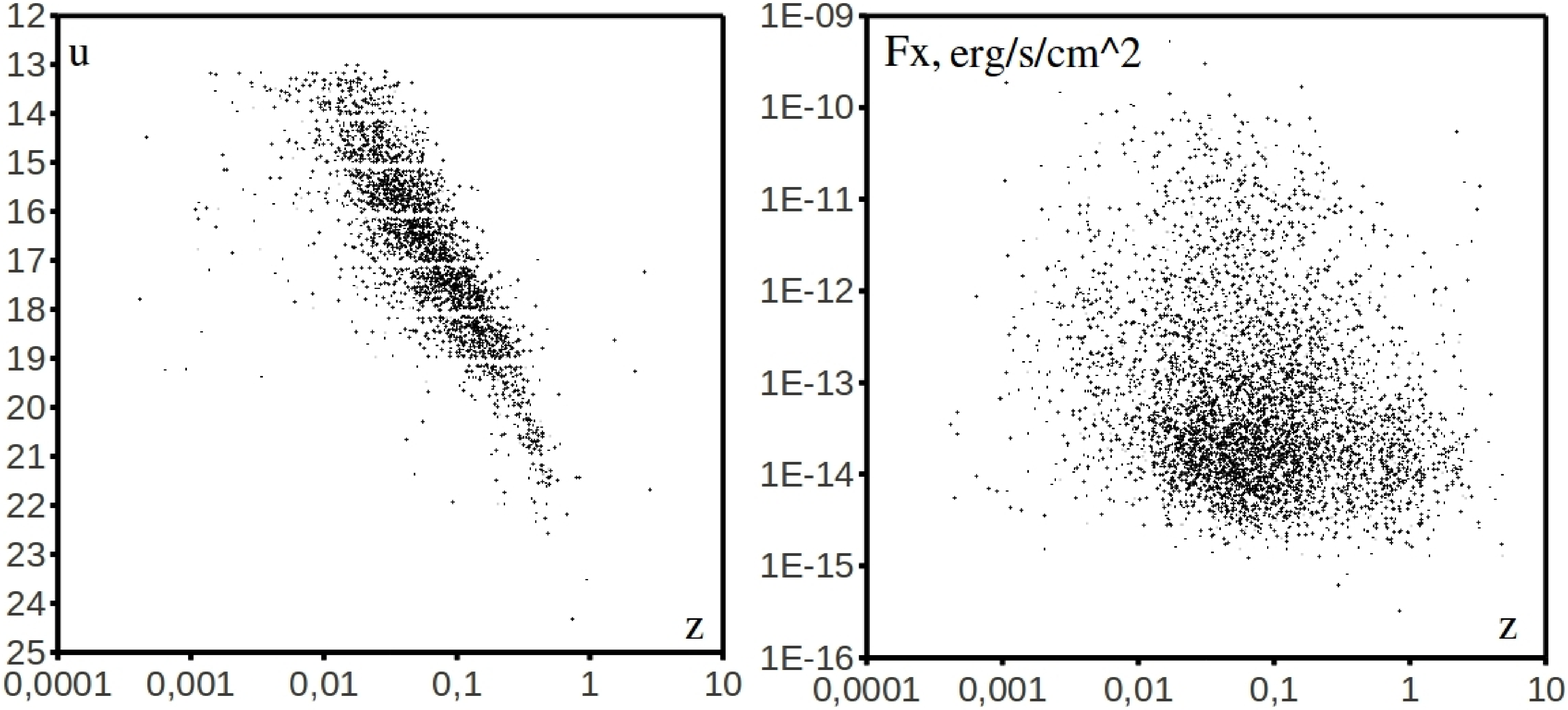} 
 \caption{Distribution of u magnitude and X-ray flux from redshift for extragalactic X-ray sources. Note the absence of dependence of X-ray flux from distance.}
   \label{fig1}
\end{center}
\end{figure}

\begin{figure}[b]
\begin{center}
 \includegraphics[width=3.4in]{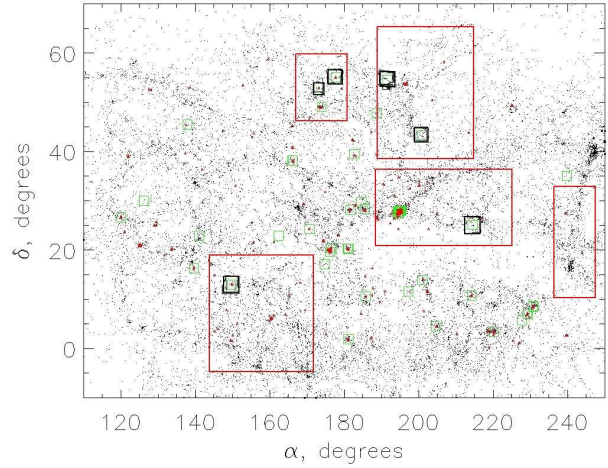} 
 \caption{Sky distribution of SDSS galaxies with radial velocities between 4000 and 11000 km/s. Larger dots are faint X-ray galaxies, squares - bright X-ray galaxies with $F_{X}>3.7\cdot 10^{-13} erg/cm^2$, rectangles - filaments from \cite{smith12}, double squares - galaxies in filaments which were considered in \cite{tugay13}.}
   \label{fig2}
\end{center}
\end{figure}

\begin{figure}[b]
\begin{center}
 \includegraphics[width=3.4in]{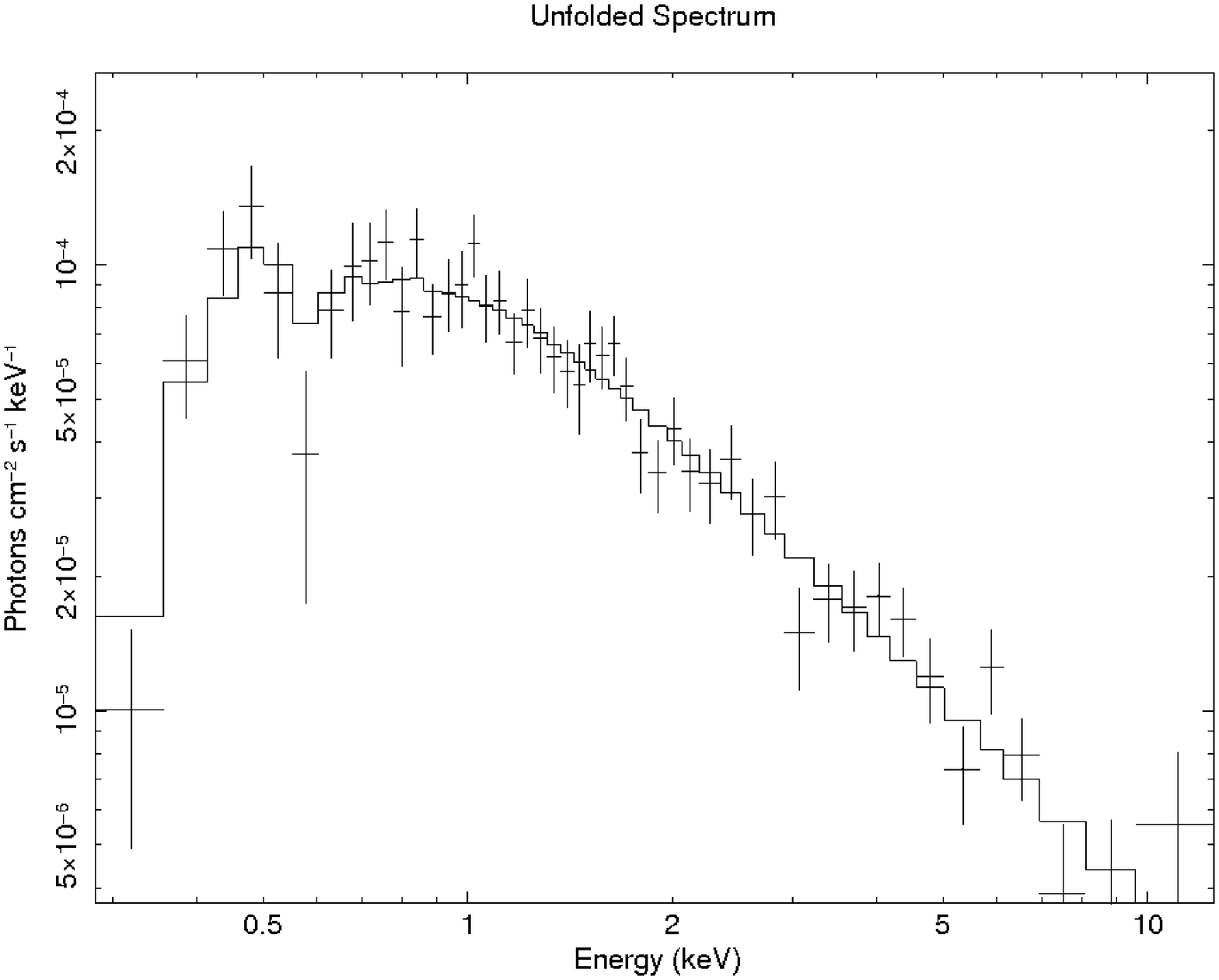} 
 \caption{Example of new X-ray spectrum of Seyfert galaxy in SDSS filament: 2MASX J15585579+0248338. Details see in \cite{tugay13a}.}
   \label{fig3}
\end{center}
\end{figure}

\section{New results}

Xgal sample was inspected to check entries of identification of more than one optical galaxy with the same X-ray source. 344 such entries were found in the whole sample. Twelve X-ray sources corresponds to 3 optical galaxies, three to 4, two to 5 and one to 6. The cases of multiple identifications in 4000-39000 km/s range includes 12 X-ray galaxy clusters, 3 groups and 8 pairs. 
392 extragalactic X-ray sources were found in Xgal sample in the same redshift range but outside SDSS sky region. These are 138 galaxy clusters, 36 groups, 3 pairs, 175 AGNs, 1 emission line galaxy, 5 irregular galaxies and 34 indefinite galaxies. AGNs includes 86 Seyfert 1 galaxies, 49 Seyfert 2 galaxies, 8 quasars, 13 blasars, 7 radio galaxies and 3 LINERs. 
It was assumed that complete filament distribution can be recovered by applying the method of steepest gradient to smoothed (sky) galaxy distribution in 100 Mpc thick slice. The results are unappropriate yet. 

\section{Conclusions}

There are enought archival data for X-ray galaxies in filaments. New methods for correct filament detection are needed. Possible directions of further study of Xgal sample includes the analysis of X-ray spectra outside SDSS region and at $z>0.1$, the description of numerous faint sources and the development of physical model of AGN X-ray spectrum.

\end{document}